Research Article

Liying Chen, Ashley P. Fidler, Alexander M. McKillop, and Marissa L. Weichman*

# Exploring the impact of vibrational cavity coupling strength on ultrafast CN + $c$-C$_6$H$_{12}$ reaction dynamics

**Abstract:** Molecular polaritons, hybrid light-matter states resulting from strong cavity coupling of optical transitions, may provide a new route to guide chemical reactions. However, demonstrations of cavity-modified reactivity in clean benchmark systems are still needed to clarify the mechanisms and scope of polariton chemistry. Here, we use transient absorption to observe the ultrafast dynamics of CN radicals interacting with a cyclohexane ($c$-C$_6$H$_{12}$) and chloroform (CHCl$_3$) solvent mixture under vibrational strong coupling of a C–H stretching mode of $c$-C$_6$H$_{12}$. By modulating the $c$-C$_6$H$_{12}$:CHCl$_3$ ratio, we explore how solvent complexation and hydrogen (H)-abstraction processes proceed under collective cavity coupling strengths ranging from 55–85 cm$^{-1}$. Reaction rates remain unchanged for all extracavity, on-resonance, and off-resonance cavity coupling conditions, regardless of coupling strength. These results suggest that insufficient vibrational cavity coupling strength may not be the determining factor for the negligible cavity effects observed previously in H-abstraction reactions of CN with CHCl$_3$.

## 1 Introduction

Manipulating chemical reactions using electromagnetic fields is a longstanding goal in physical chemistry [1, 2], as evidenced in the rich histories of coherent control and vibrational mode-selective chemistry [2–6]. However, achieving efficient photonic control of solution-phase chemical reactions has proven challenging [7–9]. The difficulty arises from rapid energy relaxation, particularly intramolecular vibrational energy redistribution (IVR), that outcompetes most desired optically-driven processes [9–11]. It is an ongoing aim to develop new photonic methods that can circumvent IVR and robustly influence solution-phase chemical reaction dynamics.

Polariton chemistry represents one emerging possibility for photonic control of complex chemistry, harnessing strong interactions with the electromagnetic field of an optical cavity rather than a laser pulse [12–18]. Polaritons, states with mixed light-matter character, are formed by bringing a confined photonic mode, often a longitudinal mode of a Fabry-Pérot optical cavity, into resonance with a sufficiently optically bright transition of a molecular ensemble (Fig. 1A). As described within the quantum optics formalism, strong coupling of $N$ molecules to a single photonic mode results in the formation of the upper and lower polariton eigenstates separated by the collective Rabi splitting, $\hbar\Omega_R \propto \mu_{eg}\sqrt{N/V}$, where $\mu_{eg}$ is the transition dipole moment for the coupled transition and $V$ is the cavity mode volume [17, 19–24]. The polaritons form alongside $N - 1$ uncoupled dark states.

A chief aim in polariton chemistry is to determine whether strongly cavity-coupled molecules can exhibit reactive behavior distinct from uncoupled free-space molecules. Recent experimental studies have reported modification of solution-phase chemical reaction rates under strong coupling of reactant vibrational modes inside microfluidic cavities [18, 25–29]. Though not without controversy [30, 31], these demonstrations have spurred theoretical efforts to unravel precisely how vibrational strong coupling (VSC) might be used to rationally modulate chemical reactivity [17, 32, 33]. At present, the precise mechanisms driving vibrational cavity-altered chemistry remain elusive, though one increasingly common proposal is that polariton states may feature distinct IVR dynamics from uncoupled molecules [32, 34–39]. These predictions await further validation in benchmark experimental systems.

Here, we investigate how VSC impacts the solution-phase reactions of cyano (CN) radicals proceeding with cyclohexane ($c$-C$_6$H$_{12}$) in chloroform solvent (CHCl$_3$). A major focus of our work is the exothermic, nearly barrierless hydrogen (H)-abstraction reaction (Fig. 1B):

$$\text{CN} + c\text{-C}_6\text{H}_{12} \rightarrow \text{HCN} + c\text{-C}_6\text{H}_{11} \qquad (1)$$

H-abstraction reactions represent a powerful platform to test predictions for vibrational polariton chemistry: they have well-established mechanisms that allow robust theoretical modelling [40, 41] and direct measurements of state-specific dynamics are feasible on ultrafast timescales [42–51]. We generate CN radicals *in situ* from ultrafast ultraviolet (UV) photolysis of ICN precursor molecules and use white light transient absorption (TA) spectroscopy to track the subsequent reactive dynamics – including geminate and diffusive photofragment recombination, solvent complexation, solvent cage escape, and H-abstraction – via their electronic absorption signatures. The rapid timescales associated with these processes allow us to follow the CN + $c$-C$_6$H$_{12}$ reaction to completion on picosecond to nanosecond timescales. By performing experiments inside optical microcavities, we can follow intracavity reaction rates under VSC of a bright C–H stretching mode of the $c$-C$_6$H$_{12}$ reactant.

This work builds on our prior study of the reaction of CN with CHCl$_3$ under VSC of the CHCl$_3$ C–H stretch [52], where we observed no cavity dependence of any time constants. One hypothesis was that no cavity effects appeared for CN + CHCl$_3$ because the $\hbar\Omega_R \sim 25$ cm$^{-1}$ Rabi splitting achieved in neat CHCl$_3$ was too small to modify chemistry. In contrast, considerably larger absolute Rabi splittings are accessible in $c$-C$_6$H$_{12}$ due to its larger infrared (IR) transition dipoles. Here, we use systematic dilutions of $c$-C$_6$H$_{12}$ in CHCl$_3$ to tune the collective cavity coupling strength from $\hbar\Omega_R \sim 55$ to 85 cm$^{-1}$ for the strongest IR-active C–H stretching mode of $c$-C$_6$H$_{12}$. These coupling values are more comparable to the $\hbar\Omega_R > 70$ cm$^{-1}$ splittings seen in literature reporting rate modifications under VSC [18, 25, 29, 34].

In this report, we detail reaction rates for the CN + $c$-C$_6$H$_{12}$ system in CHCl$_3$ under extracavity, on-resonance, and off-resonance conditions over a range of collective coupling strengths. As in our previous work [52], we employ dichroic microfluidic optical cavities to directly capture the dynamical signatures of CN radicals at visible wavelengths while simultaneously achieving VSC of their $c$-C$_6$H$_{12}$ co-reactants (Fig. 1C). We observe no cavity-dependent dynamics in this system for any conditions explored herein. Our null results help shed light on which properties may make

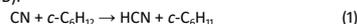

*Corresponding author: **Marissa L. Weichman**, Department of Chemistry, Princeton University, Princeton, NJ, USA. E-mail: weichman@princeton.edu; https://orcid.org/0000-0002-2551-9146
**Liying Chen, Ashley P. Fidler, and Alexander M. McKillop**, Department of Chemistry, Princeton University, Princeton, NJ, USA.

certain reactions susceptible to modulation with VSC. In particular, we argue that the lack of cavity effects for this low-barrier, exothermic system is not due to insufficient $\hbar\Omega_R$, but rather is consistent with recent literature proposals that vibrational polariton chemistry may arise from cavity-mediated IVR.

## 2 Materials and Methods

Our ultrafast white light (WL) transient absorption spectroscopy setup is described in detail in our previous report [52]. Briefly, we use a 7 mJ/pulse Ti:sapphire laser (Astrella, Coherent), that produces 800 nm, 60 fs pulses at a 1 kHz repetition rate to generate UV pump, visible WL probe, and mid-IR monitor beams. We use an optical parametric amplifier (OPerA Solo OPA FH/SHSF, Light Conversion) to generate pump pulses with a central wavelength of 250 nm. We generate our WL supercontinuum probe pulses by focusing 800 nm light into a 3 mm thick CaF$_2$ crystal [53]. We introduce a delay in the probe beam relative to the pump using a motorized delay line (DL325, Newport) with a total travel distance of 325 mm (2.2 ns). We set the polarization of the pump beam to the magic angle (54.7°) relative to the probe beam for all measurements. The UV pump and WL probe are focused into the sample cell with a crossing angle of ~7°. We propagate the WL probe into a grating spectrometer and CCD camera (Isoplane-320 and Blaze-400B, Princeton Instruments). We monitor cavity transmission before and after each TA experiment using mid-IR light generated with a second optical parametric amplifier (OPerA Solo OPA NDFG, Light Conversion). The mid-IR pulses are centered at 3400 nm (2940 cm$^{-1}$) with a full-width at half-maximum (FWHM) bandwidth of ~250 cm$^{-1}$. The mid-IR output propagates collinearly with the WL probe and hits the sample at normal incidence. The pump, probe, and IR monitor beams have respective diameters of approximately 1050 μm, 200 μm, and 300 μm at the sample. We record the mid-IR spectrum using a diffraction grating and array detector (2DMCT, PhaseTech). We acquire all spectra using home-written LabVIEW programs and process data in MATLAB and R Studio.

We prepare 0.3 M ICN sample solutions by dissolving ICN (98% purity, Thermo Scientific Chemicals) into mixtures of cyclohexane (99.7% purity, Sigma-Aldrich) and chloroform (99.8% purity, Sigma-Aldrich). We work with five different solvent mixtures with *c*-C$_6$H$_{12}$:CHCl$_3$ volume ratios spanning 2:20 to 8:20, which correspond to *c*-C$_6$H$_{12}$ molarities of 0.84 to 2.64 M. Our upper limit for the *c*-C$_6$H$_{12}$ concentration is determined by the limited solubility of ICN in *c*-C$_6$H$_{12}$; this also prevents us from conducting experiments in neat *c*-C$_6$H$_{12}$.

We deliver sample solutions to a microfluidic demountable flow cell (TFC-M13-3, Harrick Scientific) with a peristaltic pump. We continuously flow solution through the microcavity so that a fresh sample is provided for each TA shot. For extracavity measurements, we fit the cell with IR-transparent CaF$_2$ windows. We assemble Fabry-Pérot cavities for VSC by fitting the cell with distributed Bragg reflector (DBR) mirrors with high reflectivity from 3000 – 3500 nm (~2800 – 3300 cm$^{-1}$, UltraFast Innovations). The poor IR reflectivity outside of this range ensures we have no spurious coupling outside of the C–H stretching region. These mirrors are also designed to maximize transparency at UV and visible wavelengths in order to transmit our TA pulses. We report the single-mirror reflectivity and empty two-mirror cavity transmission spectra for both IR and visible wavelengths in Section S1 of the supplementary material.

We employ a 56 μm polytetrafluoroethylene (PTFE) spacer between the DBR mirrors to achieve both a workable path length for TA and a reasonable cavity free spectral range of ~100 cm$^{-1}$. Depending on the parallelism of a given cavity assembly, translating the cavity by 3.5 mm in the plane of the mirrors typically moves the cavity fringes by ~35 cm$^{-1}$, corresponding to a change in cavity length of ~0.6 μm. We exploit this imperfect parallelism to probe distinct cavity resonance conditions simply by translating the cavity assembly relative to the laser beams. Spatial variation in the cavity length over the IR monitor beam diameter serves to slightly broaden the observed cavity transmission fringes, which range from ~6 cm$^{-1}$ FWHM for an empty cavity to ~20 cm$^{-1}$ FWHM when filled with sample solution. Since the IR monitor beam spot is slightly larger than that of the WL probe, this broadening reports on the range of cavity-coupling conditions experienced within the TA measurement volume.

## 3 Results

### 3.1 Vibrational strong coupling in *c*-C$_6$H$_{12}$

We first demonstrate that we can reach the VSC regime in *c*-C$_6$H$_{12}$ and modulate the collective cavity coupling strength via dilution in CHCl$_3$. *c*-C$_6$H$_{12}$ has four strong IR-active C–H stretching modes, two of $a_{2u}$ symmetry and two of $e_u$ symmetry. Due to anharmonicity and Fermi resonances, these modes appear in the IR spectrum as two bands centered near 2850 cm$^{-1}$ and 2930 cm$^{-1}$ [54]. Here we target the stronger of these two bands which, under our solution-phase conditions, appears centered at 2927 cm$^{-1}$ with a linewidth of ~11 cm$^{-1}$ FWHM (black trace in Fig. 2A). To establish VSC, we inject a *c*-C$_6$H$_{12}$:CHCl$_3$ solvent mixture into our microfluidic cavity fitted with DBR mirrors and compress the cavity length until one optical mode is on resonance with the target transition. In $L = 56$ μm cavities, we typically couple to longitudinal modes of order $n \sim 40 - 50$. Despite spectral congestion in the C–H stretching region, we are able to identify an on-resonance coupling condition where only the target mode of *c*-C$_6$H$_{12}$ at 2927 cm$^{-1}$ is strongly coupled, while the neighboring 2850 cm$^{-1}$ band of *c*-C$_6$H$_{12}$ and the 3020 cm$^{-1}$ mode of CHCl$_3$ are reasonably detuned from resonance with their nearest cavity modes (see Section S2 of the supplementary material). The resulting cavity transmission spectra feature Rabi splittings from $\hbar\Omega_R \sim 55$ to 85 cm$^{-1}$ as we increase the cyclohexane concentration in CHCl$_3$ from 0.84 M to 2.64 M (Fig. 2A). Rabi splittings for each concentration are listed in Table S1 of the supplementary material. For each concentration, the Rabi splitting significantly exceeds both the molecular C–H stretching linewidth (~11 cm$^{-1}$ FWHM) and the cavity mode linewidth (~20 cm$^{-1}$ FWHM), confirming that the system is well within the VSC regime. We construct dispersion plots by measuring cavity transmission spectra as a function of cavity detuning from resonance. A representative dispersion plot for the 2.64 M *c*-C$_6$H$_{12}$ solution is provided in Section S2 of the supplementary material.

We simulate our experimental cavity transmission spectra using the classical expression for the frequency-dependent intensity of light transmitted through a Fabry-Pérot cavity containing an absorbing medium [55–57]. The simulated Rabi splittings closely match the experimental Rabi splitting for each *c*-C$_6$H$_{12}$:CHCl$_3$ mixture (Fig. 2B). To perform these simulations, we make use of the complex refractive indices of CHCl$_3$ and *c*-C$_6$H$_{12}$ made available by Pacific Northwest National Laboratories (PNNL) [58]. This procedure is described further in Section S3 of the supplementary material.

### 3.2 Transient absorption and fitting of extracavity CN + *c*-C$_6$H$_{12}$ reaction dynamics

We now detail our ultrafast extracavity measurements performed in IR-transparent microfluidic cells before moving on to our cavity-coupled results in Section 3.3. We monitor the ultrafast dynamics of CN radicals using WL TA spectroscopy following UV photolysis of ICN in *c*-C$_6$H$_{12}$:CHCl$_3$ mixtures. We analyze the resulting TA spectra using the procedures established in our prior ICN photodissociation experiments [52].

Fig. 3A presents representative TA data for 0.3 M ICN in a 2.64 M *c*-C$_6$H$_{12}$:CHCl$_3$ solution. We observe two features centered at 390 nm and 340 nm with distinct time-domain dynamics. Following previous work in related systems [49, 52], we assign the 390 nm feature to the $\tilde{B} \leftarrow \tilde{X}$ electronic transition of free, weakly-solvated CN radicals which appear rapidly following photolysis of ICN. Over time, these CN radicals form strongly-bound CN-solvent complexes and their absorption blue-shifts to 340 nm. The rise of the 340 nm feature therefore reflects the formation of CN-solvent complexes while its decay is associated with complex loss due to H-abstraction.

To discern precise time constants from the overlapping 390 and 340 nm features, we use a multivariate regression to decompose the absorption spectrum acquired at each pump-probe time delay into three spectral components. Fig. 3B shows the decomposition of the TA spectrum at 1.5 ps. We fit the 390 nm feature with a Lorentzian and the 340 nm feature with an experimental TA trace obtained 100 ps, after the 390 nm feature has decayed. We use a third regression component to capture a broad Gaussian-like feature whose linewidth exceeds the bandwidth of our probed spectral window. We dub this third component the "solvent" feature and ascribe it to a combination of the I*($^2$P$_{1/2}$)-solvent charge transfer band, two-photon solvent absorption, and absorption artifacts of the DBR mirrors [42-46, 49, 52, 59]. We capture the solvent component in our regression using an experimental TA spectrum acquired at 1000 ps, after the 340 nm feature has decayed.

We use the regression to construct time traces for the 390 and 340 nm features, as shown for the 2.64 M *c*-C$_6$H$_{12}$:CHCl$_3$ system in Figs. 3C and 3D. The 390 nm feature exhibits a single exponential decay with a <10 ps time constant, attributed to the loss of free CN through photofragment recombination, isomerization, solvent-complexation, and H-abstraction [42, 43, 45, 49, 52]. The 340 nm feature is well-described by <2 ps exponential growth followed by a few-100 ps decay, as solvent-complexed CN species are formed and then lost to H-abstraction reactions [49, 52].

### 3.3 Intracavity CN + *c*-C$_6$H$_{12}$ reaction dynamics

Extracavity time constants for the 390 and 340 nm features are shown in Fig. 4 (black circles) for all *c*-C$_6$H$_{12}$ concentrations. All time constants decrease as the relative concentration of *c*-C$_6$H$_{12}$ increases. As both *c*-C$_6$H$_{12}$ and CHCl$_3$ are present as co-solvents in our experiments, these trends reflect the relative propensities of the two solvents to mediate CN dynamics. The 390 nm decay constant exhibits the most pronounced change as the *c*-C$_6$H$_{12}$:CHCl$_3$ ratio is scanned (Fig. 4A), indicating that photofragment recombination and solvent caging dynamics are highly solvent-specific. The 340 nm decay constant also drops as the *c*-C$_6$H$_{12}$:CHCl$_3$ ratio increases (Fig. 4C). Notably, the 340 nm feature displays a ~400 ps decay even for the smallest 0.84 M *c*-C$_6$H$_{12}$ concentration, which is considerably faster than the ~1500 ps decay observed in neat CHCl$_3$ [52]. We conclude that the H-abstraction reactions of CN with *c*-C$_6$H$_{12}$ outcompete those with CHCl$_3$, even when CHCl$_3$ is present as co-solvent. We are therefore confident that our observed time constants for the *c*-C$_6$H$_{12}$:CHCl$_3$ system report largely on the CN + *c*-C$_6$H$_{12}$ reaction rate, with only minor contributions from CN + CHCl$_3$. Our findings are consistent with prior extracavity literature reports of the CN + *c*-C$_6$H$_{12}$ system in various chlorinated solvents [46-48, 50, 51].Following our characterization of the CN + *c*-C$_6$H$_{12}$ system under extracavity conditions, we now turn to examine the potential impact of vibrational cavity coupling on these reaction dynamics. We perform TA experiments for ICN in each *c*-C$_6$H$_{12}$:CHCl$_3$ solvent mixture inside DBR microcavities tuned both on-resonance with the strongest C–H stretching transition of *c*-C$_6$H$_{12}$ (see Fig. 2A) and tuned off-resonance from any strong *c*-C$_6$H$_{12}$ absorption (see Fig. S2 of the supplementary material). We extract time constants for the 390 nm and 340 nm features for each solvent mixture and each coupling condition. These intracavity results are summarized in Fig. 4 (purple and orange points). Additional data are broken down by individual TA scans in Section S4 of the supplementary material. We note that across all our data sets, the 340 nm time constants tend to show more variability than those of the 390 nm feature due to the 340 nm feature's lower signal intensity and increased spectral overlap with solvent features. In addition, our fits of intracavity data generally exhibit larger variations than extracavity data due to TA artifacts from the DBR mirrors, as discussed in our previous report [52].

Regardless, our extracted time constants show no notable differences between on-resonance, off-resonance and extracavity measurements within our temporal resolution and noise levels. We therefore find that VSC of *c*-C$_6$H$_{12}$ has a negligible impact on the recombination, isomerization, solvent-complexation, or H-abstraction rates of CN radicals in *c*-C$_6$H$_{12}$:CHCl$_3$ solvent mixtures. Moreover, we see no discrepancies arise as we sweep the collective vibrational cavity coupling strength from 55 to 85 cm$^{-1}$. Cavity coupling strength may therefore not represent the primary parameter determining whether cavity coupling modulates either the CN + *c*-C$_6$H$_{12}$ system reported here or the CN + CHCl$_3$ system we studied previously [52].

## 4 Discussion

In this work, we continue to develop our experimental framework targeting vibrational polariton chemistry in the simplest accessible solution-phase bimolecular reactions. We track the ultrafast dynamics of CN radicals interacting with a *c*-C$_6$H$_{12}$:CHCl$_3$ solvent mixture under VSC of the *c*-C$_6$H$_{12}$ C–H stretch. We adjust the cavity coupling strength by modulating the *c*-C$_6$H$_{12}$ concentration. In doing so, we examine whether vibrational cavity coupling strength is a determining factor for the negligible cavity effects we have observed in H-abstraction reactions of CN with both *c*-C$_6$H$_{12}$ and CHCl$_3$ under VSC [52]. In our previous study of CN + CHCl$_3$, we speculated that the small collective Rabi splitting of only $\hbar\Omega_R \sim 25$ cm$^{-1}$ was insufficient to significantly impact reaction rates. Here, we obtain similar null cavity-modification results in the CN + *c*-C$_6$H$_{12}$ system despite reaching collective Rabi splittings more than three times larger than is possible in CHCl$_3$. The Rabi splittings achieved here are the same order of magnitude as those accessed in compelling reports of vibrational polariton chemistry [18, 25, 29, 34]. We therefore conclude that insufficient vibrational coupling strength is not the sole factor influencing whether simple bimolecular H-abstraction reaction rates can be modulated with cavity coupling.

We now consider other potential explanations for the observation of negligible cavity modification of CN + *c*-C$_6$H$_{12}$, building on the more extensive discussion in our prior report [52]. One possibility is that our TA signals are dominated by the reactions of uncoupled molecules – molecules located at nodes of the cavity field or whose transition dipoles lie out of the plane of the cavity mirrors. Another possibility is that we are performing strong coupling with much higher cavity mode orders ($n \sim 40 - 50$) than are used in much of the VSC literature. While the impact of cavity mode order on polariton chemistry is not yet well understood, it may be that our cavity mode volumes are too large – or our per-molecule coupling strengths too small – to observe cavity alteration of chemistry. Unfortunately, we have thus far found it difficult to extract clear TA signals in thinner, lower-order cavities [52]. Pursuing these weak signals may be the subject of future work.

A final, more interesting, possibility for our null results is that – as several current theories and some experimental evidence have begun to suggest [34, 35, 37-39] – vibrational polariton chemistry is driven by cavity-modified IVR dynamics [32, 34-39]. We do not expect IVR to impact the dynamics of the near barrierless, highly exothermic H-abstraction reactions studied here, which should proceed rapidly regardless of the level of vibrational excitation of reactants. Our null results are therefore consistent with an IVR-related mechanism for vibrational polariton chemistry.

In future work, we will consider cavity-modification of bimolecular reaction systems that require the injection of vibrational energy to proceed. By targeting IR-driven solution-phase reactions, such as the endothermic H-abstraction reactions of bromine atoms with various cavity-coupled solvents [8], we can optically pump the vibrational solvent polaritons and determine how their reactive trajectories differ from uncoupled vibrationally-excited states. These ideas have already been applied to great effect in 2D-IR studies of unimolecular vibrational polariton dynamics [34]. By performing benchmark experiments on simple, well-understood systems we hope to further clarify the scope and prospects for vibrational cavity control of chemical reactivity.

## Authors' statements

**Supplementary Material:** See the supplementary material for mirror characterization, experimental Rabi splittings of each solvent system, cavity dispersion plots, description of classical cavity transmission simulations, and more detailed tables of extracted ultrafast time constants.

**Acknowledgements:** We thank Tanya Myers for providing quantitative frequency-dependent complex refractive index data for chloroform and cyclohexane measured by Pacific Northwest National Laboratory and Intelligence Advanced Research Projects Activity.

**Research funding:** This work was supported by the US Department of Energy, Office of Science, Basic Energy Sciences, CPIMS Program under Early Career Research Program award DE-SC0022948, by the Princeton Catalysis Initiative, and by Princeton University start-up funds. We also acknowledge funds from the Princeton Center for Complex Materials (PCCM) National Science Foundation (NSF) Materials Research Science and Engineering Center (MRSEC; DMR-2011750), which supported the construction of our infrared spectrometer. APF acknowledges support from a Princeton Presidential Postdoctoral Research Fellowship.

**Author contribution:** All authors have accepted responsibility for the entire content of this manuscript and approved its submission.

**Conflict of interest:** The authors state no conflicts of interest.

**Informed consent:** Informed consent was obtained from all individuals included in this study.

**Ethical approval:** The conducted research is not related to either human or animal use.

**Data availability statement:** The datasets generated and/or analysed during the current study are available from the corresponding author upon reasonable request.

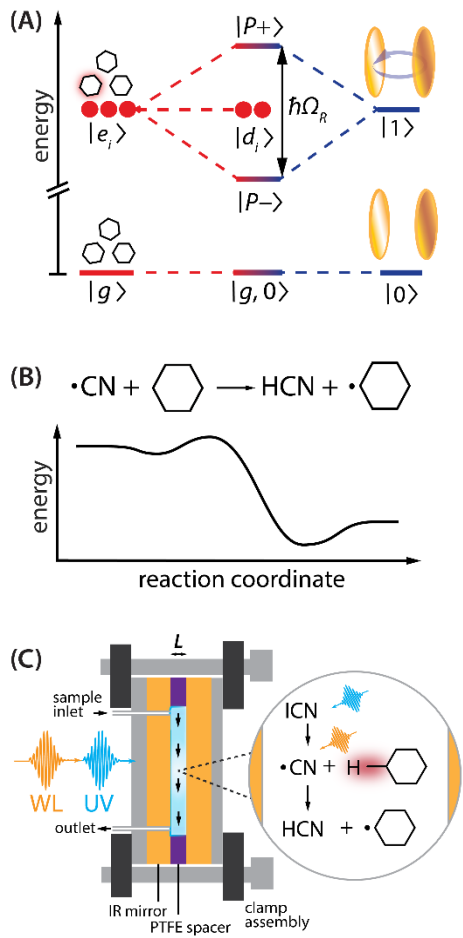

**Fig. 1: (A)** Energy diagram of collective vibrational strong coupling as illustrated within the Tavis-Cummings model. When the photonic excitation of a quantized cavity mode ($|0\rangle \rightarrow |1\rangle$) is resonantly matched to a molecular transition ($|g\rangle \rightarrow |e_i\rangle$), the system hybridizes to produce the upper and lower polariton states ($|P-\rangle, |P+\rangle$), separated by the Rabi splitting ($\hbar\Omega_R$), alongside a manifold of dark states with molecular character ($|d_i\rangle$). **(B)** Schematic energy diagram for the exothermic, nearly barrierless CN + $c$-$C_6H_{12}$ reaction. **(C)** Scheme for ultrafast studies of intracavity H-abstraction reactions of CN with cavity-coupled $c$-$C_6H_{12}$ following photolysis of ICN. Experiments are performed in a dichroic Fabry-Pérot cavity housed within a microfluidic flow cell. We probe intracavity reactions using ultraviolet (UV) pump – white light (WL) probe transient absorption spectroscopy.

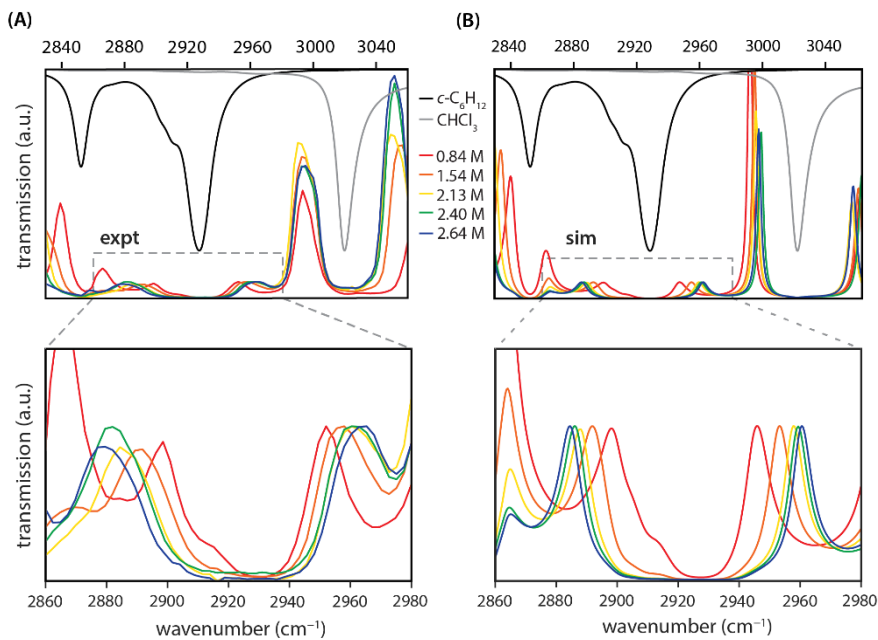

**Fig. 2: (A)** Experimental and **(B)** simulated cavity transmission spectra for various $c$-$C_6H_{12}$:$CHCl_3$ mixtures in distributed Bragg reflector (DBR) Fabry-Pérot cavity fitted with an $L$ = 56 μm cavity spacer. Colored traces show transmission spectra for on-resonance cavities filled with 0.84 M (red), 1.54 M (orange), 2.13 M (yellow), 2.40 M (green), and 2.64 M (blue) solutions of $c$-$C_6H_{12}$ in $CHCl_3$. Rabi splittings are evident in all traces and increase with $c$-$C_6H_{12}$ concentration. The absorption spectra of $c$-$C_6H_{12}$ and $CHCl_3$ are plotted in black and grey, respectively, using reference data from PNNL [58]. The zoomed-in bottom panels better illustrate the Rabi splittings arising from strong coupling of the brightest $c$-$C_6H_{12}$ C–H stretching mode centered at 2927 cm$^{-1}$.

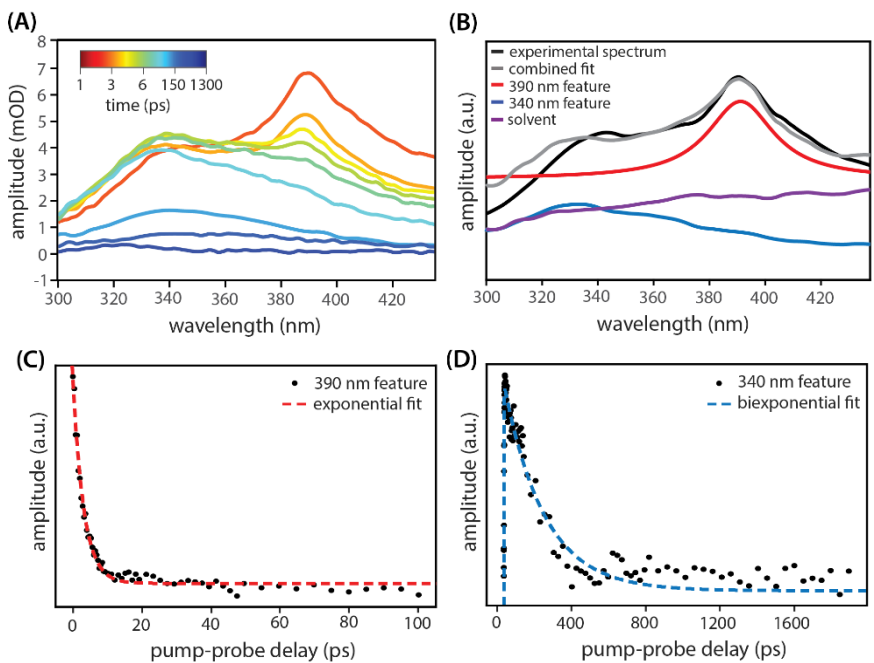

**Fig. 3:** Extracavity transient absorption (TA) dynamics of CN radicals produced following photolysis of ICN in a 2.64 M solution of $c$-$C_6H_{12}$ in $CHCl_3$. **(A)** TA spectra obtained at pump-probe time delays from 1 to 1300 ps. The free CN feature at 390 nm decays as the CN-solvent complex feature at 340 nm grows in and subsequently decays. **(B)** Decomposition of a TA spectrum collected at a 1.5 ps time delay (black). We model the 390 nm component (red) as a Lorentzian function. We derive the profile of the component at 340 nm (blue) from a TA spectrum collected 100 ps after ICN photolysis. We attribute the broad solvent feature (purple) to an I*($^2P_{1/2}$)-solvent charge transfer band and other secondary processes. We model this solvent component using a TA spectrum collected at 1000 ps. The regression components sum to the combined fit (grey). Note that we denoise the TA spectra plotted in panels A and B for visual clarity, but all fits are performed with raw data. **(C)** Time-dependent weighting of the 390 nm spectral component extracted from TA measurements (black dots) and single exponential fit (red). **(D)** Time-dependent weighting of the 340 nm spectral component extracted from TA measurements (black dots) and biexponential fit (blue).

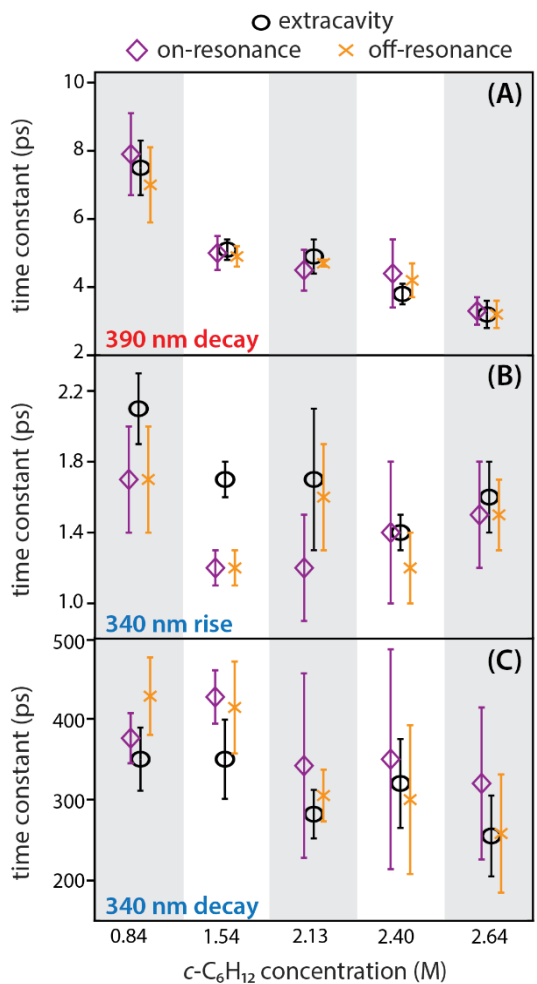

**Fig. 4.** Experimental time constants for the CN + $c$-$C_6H_{12}$ reaction system measured in cavities tuned on-resonance (purple diamonds) and off-resonance (orange crosses) with respect to the C–H stretch of $c$-$C_6H_{12}$ at 2927 cm$^{-1}$, as compared to extracavity results (black circles). Time constants are reported for each solvent system. **(A)** Time constants for decay of the free CN radical feature at 390 nm. **(B)** Time constants for the rise of the CN-solvent complex feature at 340 nm. **(C)** Time constants for the decay of the CN-solvent complex feature at 340 nm. Error bars represent standard deviations arise from averaging time constants from the individual datasets presented in Tables S2-S6.

- SUPPLEMENTARY MATERIAL -

**Exploring the impact of vibrational cavity coupling strength on ultrafast CN + *c*-C$_6$H$_{12}$ reaction dynamics**

Liying Chen,[1] Ashley P. Fidler,[1] Alexander M. McKillop,[1] and Marissa L. Weichman[1,*]

[1]Department of Chemistry, Princeton University, Princeton, New Jersey 08544, USA

*weichman@princeton.edu

**Section S1: Wavelength-dependent mirror reflectivity and transmission**

We use custom-coated distributed Bragg reflector (DBR) mirrors (UltraFast Innovations GmbH) engineered with alternating layers of high and low refractive index materials ($HfO_2$, $SiO_2$). The design allows for optical transmission of ultraviolet and visible light (Fig. S1AB) while maintaining high reflectivity in the C−H stretching region (Fig. S1CD). We assess the DBR mirror reflectivity profile using an infrared imaging microscope (Nicolet iN10 Mx, Thermo Scientific). These mirrors display a narrow band of high reflectivity (R > 90%) in the C−H stretching region spanning ~ 2800 – 3400 $cm^{-1}$ (Fig. S1C). Upon assembly into an air-filled 56 μm Fabry–Pérot cavity, we observe a cavity transmission spectrum featuring ~6 $cm^{-1}$ FWHM cavity fringes separated by a ~100 $cm^{-1}$ FSR in the C–H stretching region (Fig. S1D).

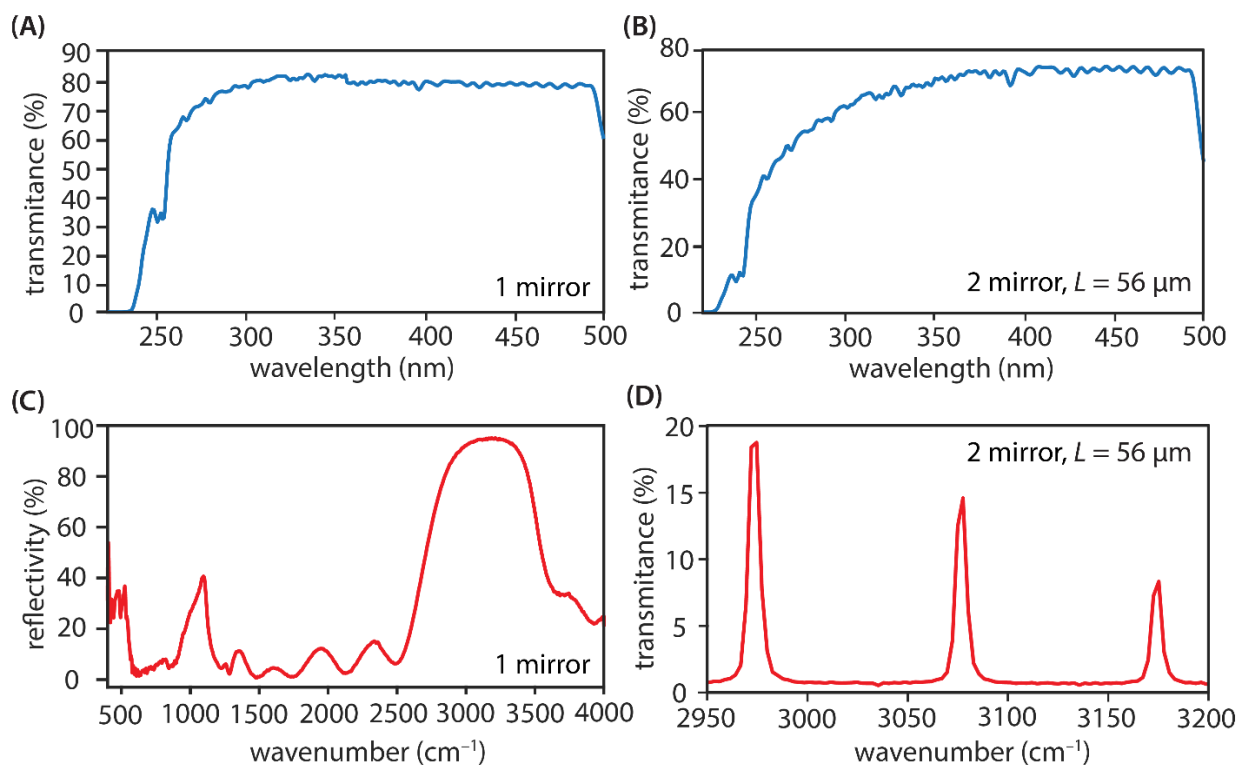

**Fig. S1**: **(A)** UV-vis transmission spectrum for a single DBR cavity mirror. **(B)** UV-vis transmission spectrum for an air-filled cavity constructed from two mirrors separated by a 56 μm spacer. **(C)** IR reflectivity spectrum of a single DBR mirror. **(D)** IR transmission spectrum of the C–H stretching region obtained for a 56 μm two-mirror cavity.

**Table S1**: Experimental Rabi splittings obtained for the *c*-$C_6H_{12}$ C–H stretching mode at 2927 cm$^{-1}$ in each *c*-$C_6H_{12}$:$CHCl_3$ solvent mixture studied here. Rabi splittings are extracted from the corresponding cavity transmission spectra in Fig. 2A of the main text.

| *c*-$C_6H_{12}$:$CHCl_3$ volume ratio | *c*-$C_6H_{12}$ molarity (M) | Rabi splitting (cm$^{-1}$) |
|---|---|---|
| 2:20 | 0.84 | 55 |
| 4:20 | 1.54 | 66 |
| 6:20 | 2.13 | 76 |
| 7:20 | 2.40 | 80 |
| 8:20 | 2.64 | 85 |

**Section S2: Experimental and simulated cavity dispersion plots for *c*-$C_6H_{12}$:$CHCl_3$ mixtures**

In Fig. S2, we compare the experimental and simulated transmission spectra for a cavity containing a 2.64 M solution of *c*-$C_6H_{12}$ in $CHCl_3$ plotted as a function of cavity length to show dispersion. We tune the cavity length by translating the assembly in the plane of the cavity and relying on the imperfect parallelism of the cavity mirrors. Experimental cavity lengths are determined by finding the best match between the experimental and simulated dispersion curves. Disparities between the experimental and simulated curves are chiefly due to broadening of the experimental cavity transmission features. We attribute this broadening to the finite resolution of our spectrometer and variations in cavity length within the experimental IR focal volume.

Both experimental and simulated dispersion curves display avoided crossing features, indicative of vibrational strong coupling (VSC) of various C−H stretching modes of both *c*-$C_6H_{12}$ and $CHCl_3$. For the on-resonance cavity-coupling conditions discussed throughout this manuscript, we target VSC of the brightest C-H stretching mode of *c*-$C_6H_{12}$ centered at 2927 cm$^{-1}$ in order to attain the largest Rabi splitting. This on-resonance coupling condition is indicated by the red vertical lines in Fig. S2. Given the spectral congestion inherent in the C−H stretching region of this system, it is challenging to identify a clear off-resonance condition where no cavity mode is coupled to any C−H stretching feature. Here, we use the condition marked by the orange vertical lines in Fig S2 as our off-resonance configuration. We emphasize that this condition may still involve some detuned coupling to the neighboring C−H stretching mode of $CHCl_3$. However, coupling of $CHCl_3$ is not expected to impact any reaction rates, as found in our previous report [1].

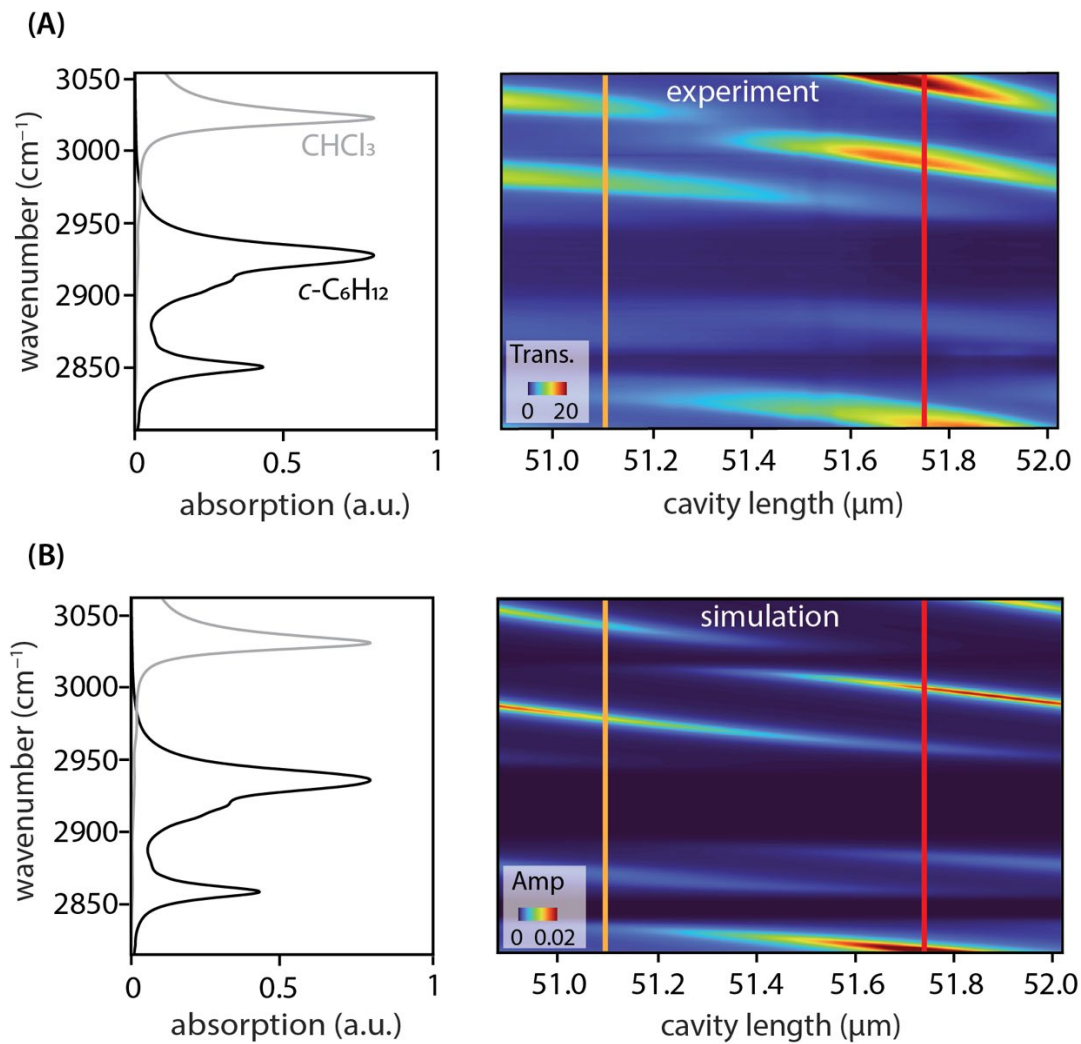

**Fig. S2**: **(A)** Experimental and **(B)** simulated DBR Fabry-Pérot cavity transmission spectra showing dispersion as a function of cavity length for the 2.64 M *c*-$C_6H_{12}$ solution in $CHCl_3$. The absorption spectra of both *c*-$C_6H_{12}$ (black) and $CHCl_3$ (grey) are plotted in the left-hand panels using reference data from PNNL [2]. The vertical lines drawn on the dispersion plots in the right-hand panels illustrate the on-resonance (red) and off-resonance (orange) conditions.

## Section S3: Classical simulation of cavity transmission spectra

To simulate experimental cavity transmission spectra, we use the classical expression for light transmitted through a two-mirror Fabry-Pérot optical cavity [3-5]:

$$\frac{I_T(\nu)}{I_0} = \frac{T(\nu)^2 e^{-\alpha(\nu)L}}{1+R(\nu)^2 e^{-2\alpha(\nu)L}-2R(\nu)e^{-\alpha(\nu)L}\cos\left(\frac{4\pi L n(\nu)\nu}{c}\right)} \quad (1)$$

Here, $T(\nu)$ and $R(\nu)$ are the frequency-dependent transmittance and reflectance for a single cavity mirror, $\alpha(\nu)$ and $n(\nu)$ are the absorption coefficient and real component of the refractive index for the intracavity material, and $L$ is the cavity length. Reference data for the real ($n(\nu)$) and imaginary ($k(\nu)$) components of the complex refractive indices of $c$-$C_6H_{12}$ and $CHCl_3$ were provided by Pacific Northwest National Laboratory (PNNL) [2]. We convert $k(\nu)$ to absorption coefficient $\alpha(\nu)$ using:

$$\alpha(\nu) = 4\pi k(\nu)\nu \quad (2)$$

## Section S4: Reproducibility of intracavity and extracavity transient absorption measurements

Ensuring the robustness of intracavity experimental results is crucial in the context of the well-documented reproducibility challenges in the field of polariton chemistry. We collected multiple independent transient absorption (TA) datasets for each cavity-coupling condition and for each $c$-$C_6H_{12}$:$CHCl_3$ solvent mixture. We extract time constants from each TA dataset following the fitting methods outlined in the main text. We present all time constants for individual TA scans in 0.84 M, 1.54 M, 2.13 M, 2.40 M, and 2.64 M solutions of $c$-$C_6H_{12}$ in Tables S2-S6.

In processing the datasets summarized in Tables S2-S6, we discard time constants based on two criteria. First, a time constant is discarded if the regression p-value exceeds 0.05, indicating poor spectral fitting during the regression. Secondly, we discard any time constants that fall more than three standard deviations from the mean for a specific cavity condition and $c$-$C_6H_{12}$ concentration. In some datasets, specific time constants could not be clearly resolved and were consequently excluded using the mentioned criteria. Given that the multivariate regression method treats each input component independently, we retained the other time constants in these datasets.

**Table S2**: Time constants obtained for the CN + *c*-$C_6H_{12}$ reaction in an 0.84 M solution of *c*-$C_6H_{12}$ in $CHCl_3$ under extracavity, on-resonance, and off-resonance conditions.

| 0.84 M *c*-$C_6H_{12}$ | 390 nm | 340 nm | |
|---|---|---|---|
| | $\tau_1$ (ps) | $\tau_1$ (ps) | $\tau_2$ (ps) |
| *extracavity* | | | |
| 2023-06-14 #1 | 7.9 | 2.2 | 360 |
| 2023-06-14 #2 | 8.7 | 2.1 | 360 |
| 2023-06-14 #3 | 7.2 | 2.4 | 410 |
| 2023-06-14 #4 | 6.5 | 2.1 | 320 |
| 2023-06-14 #5 | 7.4 | 1.8 | 300 |
| **Mean** | **7.5 ± 0.8** | **2.1 ± 0.2** | **350 ± 40** |
| *on-resonance* | | | |
| 2023-06-09 #1 | 7.3 | 2.0 | 410 |
| 2023-06-09 #2 | 8.1 | 1.6 | 320 |
| 2023-06-09 #3 | 9.6 | 1.3 | 380 |
| 2023-06-09 #4 | 8.4 | 1.7 | 390 |
| 2023-06-09 #5 | 6.2 | 1.9 | 380 |
| **Mean** | **7.9 ± 1.2** | **1.7 ± 0.3** | **380 ± 30** |
| *off-resonance* | | | |
| 2023-06-09 #1 | 6.5 | 1.5 | 460 |
| 2023-06-09 #2 | 8.1 | 2.0 | 430 |
| 2023-06-09 #3 | 6.9 | 1.5 | 480 |
| 2023-06-09 #4 | 8.0 | 1.7 | 410 |
| 2023-06-09 #5 | 5.6 | 2.0 | 360 |
| **Mean** | **7.0 ± 1.1** | **1.7 ± 0.3** | **430 ± 50** |

**Table S3**: Time constants obtained for the CN + *c*-C$_6$H$_{12}$ reaction in a 1.54 M solution of *c*-C$_6$H$_{12}$ in CHCl$_3$ under extracavity, on-resonance, and off-resonance conditions. Missing time constants were discarded due to poor fitting, as described in Section S3.

| 1.54 M *c*-C$_6$H$_{12}$ | **390 nm** $\tau_1$ (ps) | **340 nm** $\tau_1$ (ps) | $\tau_2$ (ps) |
|---|---|---|---|
| *extracavity* | | | |
| 2023-06-16 #1 | 5.5 | 1.78 | 390 |
| 2023-06-16 #2 | 4.8 | 1.50 | 340 |
| 2023-06-16 #3 | 5.0 | 1.62 | 280 |
| 2023-06-16 #4 | 5.3 | 1.67 | 340 |
| 2023-06-16 #5 | 5.0 | 1.76 | 410 |
| **Mean** | **5.1 ± 0.3** | **1.67 ± 0.12** | **350 ± 50** |
| *on-resonance* | | | |
| 2023-06-13 #1 | 4.4 | 1.15 | 400 |
| 2023-06-13 #2 | 5.3 | | |
| 2023-06-13 #3 | 4.5 | 1.11 | 450 |
| 2023-06-13 #4 | 5.3 | 1.33 | 460 |
| 2023-06-13 #5 | 5.6 | 1.31 | 400 |
| **Mean** | **5.0 ± 0.5** | **1.23 ± 0.11** | **430 ± 30** |
| *off-resonance* | | | |
| 2023-06-13 #1 | 5.2 | 1.13 | 400 |
| 2023-06-13 #2 | 4.8 | 1.30 | 390 |
| 2023-06-13 #3 | 4.4 | 1.41 | 400 |
| 2023-06-13 #4 | 4.9 | 1.30 | 510 |
| 2023-06-13 #5 | 5.1 | 1.05 | 370 |
| **Mean** | **4.9 ± 0.3** | **1.24 ± 0.15** | **410 ± 60** |

**Table S4**: Time constants obtained for the CN + $c$-$C_6H_{12}$ reaction in a 2.13 M solution of $c$-$C_6H_{12}$ in $CHCl_3$ under extracavity, on-resonance, and off-resonance conditions. Missing time constants were discarded due to poor fitting, as described in Section S3.

| 2.13 M $c$-$C_6H_{12}$ | 390 nm $\tau_1$ (ps) | 340 nm $\tau_1$ (ps) | 340 nm $\tau_2$ (ps) |
|---|---|---|---|
| *extracavity* | | | |
| 2023-06-20 #1 | 5.7 | 1.4 | 330 |
| 2023-06-20 #3 | 4.6 | 1.5 | 260 |
| 2023-06-20 #4 | 4.6 | 2.3 | 280 |
| 2023-06-20 #5 | 4.9 | 1.7 | 280 |
| **Mean** | **4.9 ± 0.5** | **1.7 ± 0.4** | **280 ± 30** |
| *on-resonance* | | | |
| 2023-06-21 #1 | 4.9 | 1.1 | 470 |
| 2023-06-21 #2 | 4.9 | 1.1 | |
| 2023-06-21 #3 | 4.9 | | |
| 2023-06-21 #4 | 3.7 | 1.0 | 250 |
| 2023-06-21 #5 | 4.1 | 1.7 | 310 |
| **Mean** | **4.5 ± 0.6** | **1.2 ± 0.3** | **340 ± 110** |
| *off-resonance* | | | |
| 2023-06-21 #1 | 4.76 | 1.3 | 260 |
| 2023-06-21 #2 | 4.82 | 1.9 | |
| 2023-06-21 #3 | 4.50 | 1.3 | 310 |
| 2023-06-21 #4 | 4.72 | 1.8 | 320 |
| 2023-06-21 #5 | 4.85 | | 330 |
| **Mean** | **4.73 ± 0.14** | **1.6 ± 0.3** | **300 ± 30** |

**Table S5**: Time constants obtained for the CN + $c$-$C_6H_{12}$ reaction in a 2.40 M solution of $c$-$C_6H_{12}$ in $CHCl_3$ under extracavity, on-resonance, and off-resonance conditions.

| 2.40 M $c$-$C_6H_{12}$ | 390 nm $\tau_1$ (ps) | 340 nm $\tau_1$ (ps) | $\tau_2$ (ps) |
|---|---|---|---|
| *extracavity* | | | |
| 2023-07-05 #2 | 3.7 | 1.4 | 250 |
| 2023-07-05 #3 | 3.2 | 1.6 | 230 |
| 2023-07-05 #5 | 4.1 | 1.5 | 290 |
| 2023-07-05 #6 | 4.3 | 1.3 | 340 |
| 2023-07-05 #7 | 3.8 | 1.2 | 370 |
| 2023-07-05 #8 | 3.6 | 1.4 | 390 |
| 2023-07-05 #9 | 3.7 | 1.5 | 350 |
| 2023-07-05 #10 | 4.1 | 1.6 | 330 |
| 2023-07-05 #11 | 4.1 | 1.5 | 340 |
| **Mean** | **3.8 ± 0.3** | **1.7 ± 0.9** | **320 ± 60** |
| *on-resonance* | | | |
| 2023-07-04 #1 | 2.7 | 1.0 | 200 |
| 2023-07-04 #2 | 4.3 | 1.5 | 560 |
| 2023-07-04 #3 | 5.8 | 2.0 | 240 |
| 2023-07-04 #4 | 4.9 | 1.1 | 280 |
| 2023-07-04 #5 | 4.0 | 1.9 | 380 |
| 2023-07-04 #6 | 4.6 | 1.0 | 440 |
| **Mean** | **4.4 ± 1.0** | **1.4 ± 0.4** | **350 ± 140** |
| *off-resonance* | | | |
| 2023-07-04 #1 | 4.6 | 1.0 | 230 |
| 2023-07-04 #2 | 4.9 | 1.0 | 430 |
| 2023-07-04 #3 | 3.6 | 1.0 | 280 |
| 2023-07-04 #4 | 4.3 | 1.5 | 170 |
| 2023-07-04 #5 | 4.0 | 1.4 | 360 |
| 2023-07-04 #6 | 4.2 | 1.3 | 320 |
| **Mean** | **4.2 ± 0.5** | **1.2 ± 0.2** | **300 ± 90** |

**Table S6**: Time constants obtained for the CN + *c*-C$_6$H$_{12}$ reaction in a 2.64 M solution of *c*-C$_6$H$_{12}$ in CHCl$_3$ under extracavity, on-resonance, and off-resonance conditions.

| 2.64 M *c*-C$_6$H$_{12}$ | 390 nm $\tau_1$ (ps) | 340 nm $\tau_1$ (ps) | 340 nm $\tau_2$ (ps) |
|---|---|---|---|
| *extracavity* | | | |
| 2023-07-11 #1 | 2.8 | 1.3 | 220 |
| 2023-07-11 #2 | 3.3 | 1.9 | 230 |
| 2023-07-11 #3 | 3.7 | 1.7 | 240 |
| 2023-07-12 #1 | 3.0 | 1.4 | 290 |
| 2023-07-12 #2 | 3.9 | 1.5 | 330 |
| 2023-07-12 #3 | 3.1 | 1.4 | 280 |
| 2023-07-12 #5 | 2.8 | 1.7 | 190 |
| **Mean** | **3.2 ± 0.4** | **1.6 ± 0.2** | **250 ± 50** |
| *on-resonance* | | | |
| 2023-07-13 #1 | 3.5 | 1.3 | 350 |
| 2023-07-14 #2 | 3.0 | 1.8 | 290 |
| 2023-07-14 #3 | 3.0 | 1.0 | 130 |
| 2023-07-18 #1 | 3.9 | 1.1 | 430 |
| 2023-07-18 #4 | 3.8 | 1.6 | 390 |
| 2023-07-18 #5 | 3.0 | 1.6 | 430 |
| 2023-07-19 #3 | 3.6 | 1.6 | 260 |
| 2023-07-19 #6 | 3.5 | 1.5 | 280 |
| 2023-07-20 #1 | 3.3 | 1.4 | 250 |
| 2023-07-20 #2 | 3.4 | 2.0 | 270 |
| 2023-07-20 #3 | 2.9 | 1.6 | 370 |
| 2023-07-20 #4 | 3.1 | 1.4 | 220 |
| 2023-07-20 #5 | 3.3 | 1.5 | 290 |
| 2023-07-20 #6 | 2.7 | 1.1 | 290 |
| 2023-07-21 #1 | 2.9 | 1.5 | 490 |
| 2023-07-21 #2 | 3.7 | 1.9 | 280 |
| 2023-07-21 #3 | 3.8 | 2.0 | 440 |
| **Mean** | **3.4 ± 0.4** | **1.5 ± 0.3** | **320 ± 90** |

*... Table S6 continued from previous page*

| 2.64 M C$_6$H$_{12}$ | 390 nm $\tau_1$ (ps) | 340 nm $\tau_1$ (ps) | $\tau_2$ (ps) |
|---|---|---|---|
| *off-resonance* | | | |
| 2023-07-14 #1 | 3.0 | 1.6 | 230 |
| 2023-07-14 #3 | 3.3 | 1.5 | 230 |
| 2023-07-18 #1 | 2.5 | 1.0 | 360 |
| 2023-07-18 #4 | 3.1 | 1.2 | 420 |
| 2023-07-18 #5 | 3.6 | 1.8 | 230 |
| 2023-07-19 #2 | 3.1 | 1.4 | 410 |
| 2023-07-19 #3 | 3.1 | 1.4 | 210 |
| 2023-07-19 #8 | 2.6 | 1.5 | 210 |
| 2023-07-20 #1 | 3.4 | 1.6 | 240 |
| 2023-07-20 #2 | 2.7 | 1.4 | 270 |
| 2023-07-20 #3 | 3.4 | 1.6 | 240 |
| 2023-07-20 #4 | 3.3 | 1.7 | 280 |
| 2023-07-20 #5 | 3.5 | 1.7 | 260 |
| 2023-07-20 #6 | 3.5 | 1.3 | 230 |
| 2023-07-21 #1 | 3.6 | 1.8 | 170 |
| 2023-07-21 #2 | 3.3 | 1.7 | 170 |
| 2023-07-21 #3 | 3.6 | 1.9 | 240 |
| **Mean** | **3.2 ± 0.4** | **1.5 ± 0.2** | **260 ± 70** |